\documentclass[journal]{IEEEtran}
\usepackage{cite}

\ifCLASSINFOpdf
   \usepackage[pdftex]{graphicx}
\else
   \usepackage[dvips]{graphicx}
\fi
\usepackage{amsmath}
\usepackage{xcolor}

\begin{document}
%
\title{Model-Free Control of Dynamical Systems with Deep Reservoir Computing}
%
%
%

\author{Daniel Canaday, Andrew Pomerance, Daniel J. Gauthier
\thanks{D.C. and D.J.G. gratefully acknowledge the support of The Ohio State University.}
\thanks{Daniel Canaday and Andrew Pomerance are with Potomac Research, LLC, Alexandria, VA 22314.  (email: daniel@potomacresear.ch) This work was done while Daniel Canaday was at the Ohio State University.}
\thanks{Daniel J. Gauthier are with the Department of Physics, Ohio State University, Columbus, OH 43210.}
}

\maketitle

\begin{abstract}
We propose and demonstrate a nonlinear control method that can be applied to unknown, complex systems where the controller is based on a type of artificial neural network known as a reservoir computer. In contrast to many modern neural-network-based control techniques, which are robust to system uncertainties but require a model nonetheless, our technique requires no prior knowledge of the system and is thus model-free. Further, our approach does not require an initial system identification step, resulting in a relatively simple and efficient learning process. Reservoir computers are well-suited to the control problem because they require small training data sets and remarkably low training times. By iteratively training and adding layers of reservoir computers to the controller, a precise and efficient control law is identified quickly. With examples on both numerical and high-speed experimental systems, we demonstrate that our approach is capable of controlling highly complex dynamical systems that display deterministic chaos to nontrivial target trajectories.
\end{abstract}


%
\IEEEpeerreviewmaketitle

\section{Introduction}

Controlling dynamical systems is a ubiquitous problem in disciplines ranging from engineering to medicine. The fundamental problem in control engineering is to design control signals that are applied to a system with accessible inputs, referred to as a plant, so that it follows a desired behavior. Solutions to this problem have far-reaching applications, such as in autonomous automobiles \cite{marino2011nested, chu2018active, fang2011robust} and aircraft \cite{stevens2016}, heating and cooling systems \cite{chiou2009application}, robotic arms \cite{ul2014design}, and chemical industrial processes \cite{nagy2007evaluation}, to name a few. 

One challenge in controlling dynamical systems is that, beyond textbook examples, they are almost always nonlinear, \textit{i.e.}, their behavior is a nonlinear function of the state variables or the accessible inputs.  Control methods for nonlinear systems include linearizing the dynamics about a typical operating point and then applying linear control methods \cite{kwakernaak1972linear}, evaluating a model of the system in real time for state estimation \cite{solitine1991}, or using artificial neural networks (ANNs) to perform the state estimation \cite{sarangapani2006}.  

Typical ANN models are feed-forward networks consisting of layers of nodes that pass the weighted outputs of nodes from previous layers through a nonlinear function  \cite{Goodfellow2016}, \textit{i.e.} the neural network forms a directed acyclic graph.  These static functions can universally represent other nonlinear functions, but they do not naturally represent time-dependent nonlinear signals. Recurrent neural networks (RNNs), in which the ANN remains directed but is no longer acyclic, naturally have temporal dynamics that can be used to learn time series, but training the recurrent weights is notoriously difficult and typically fails to converge.  This can be addressed by considering restricted RNN topologies.  A popular one is the long short-term memory architecture which has recently been applied to control \cite{hess2015}.

Alternatively, we adopt an approach based on a new paradigm for machine learning - known as reservoir computing (RC) - that is well suited for dynamical systems \cite{jaeger2001,tanaka2019recent}.  Here, the nodes of a RNN have their own dynamics and are hypothesized to be a universal dynamical system \cite{gauthier2018reservoir}.  Reservoir computing can learn the `climate' or phase-space attractor of complex systems using only a segment of the temporal evolution of a system observable \cite{lu2017reservoir}. Compared to deep learning, RC requires substantially less data to achieve good performance, requires much less training time due to the simplicity of the training algorithm, and achieves state-of-the-art performance in time-series prediction \cite{pathak2018model,vlachas2019,chattopadhyay2019,bompas2020}, system identification \cite{jaeger2003adaptive}, and spoken-word recognition \cite{larger2017high}. Because most of the network is unchanged during the training process, RC is particularly well-suited to computing with dedicated-purpose and low-power hardware, including novel implementations with delayed optical feedback \cite{larger2017high} and electronic Boolean circuits \cite{canaday2018rapid}.

As noted above, RC is well-known for its ability to form data-driven models for complex time-series. In the control context, this task can be thought of as creating a model for a plant in the absence of inputs. Unsurprisingly, this notion can be extended to include plants with accessible inputs \cite{khodabandehlou2017echo}, a task commonly referred to as \textit{system identification}.

Once a plant is identified, a control law can be devised.  Most often, closed-loop control is desired, where the plant input is a function of the desired plant observable and the actual plant observable. Techniques for obtaining such a function (or control law) are wide-ranging and include using a piece-wise linear approximation of the plant to direct construction with a feed-forward ANN; see, \textit{e.g.}, Ref.\ \cite{paraskevopoulos2017modern} for a recent review.

System identification is a common first step for controlling an unknown system, particularly when applying machine-learning based techniques such as ANNs. Recently, it has been shown \cite{antonik2016} that this two-step process is not necessary with RC. In fact, reservoir computers are capable of \textit{directly} learning an appropriate control law. This is accomplished by learning the system's `inverse,' which we explain further in the following sections.

The first contribution of this paper is to expand the study of the RC-control method introduced in Ref.~\cite{antonik2016}. We provide motivation for the algorithm, explicitly demonstrate and quantify the ability of a class of RC known as an echo state network (ESN) \cite{jaeger2001echo} to `invert' a system, and study optimizing the RC parameters that are specific to the control problem. The second contribution is to develop an iterative technique for adding parallel layers to the ESN controller, forming a deep ESN (dESN) \cite{gallicchio2017deep} to achieve more precise control. We demonstrate the efficacy of the proposed algorithm with numerical and experimental results.

The rest of this article is organized as follows. First, we define notation and formulate the control problem. We follow by explaining the concept of direct inverse control and how it can be accomplished with RC. Next, we examine the the effects of varying hyperparameters using numerical studies on controlling the Mackey-Glass chaotic system. We then develop our multi-layered control algorithm for precise control and apply this algorithm to a number of numerical and experimental examples. Finally, we conclude and discuss future research directions.

\section{Problem Formulation}\label{sec:problem_formulation}

We assume that the plant is described by 
\begin{align}
 \dot{\textbf{x}} &=\textbf{f}\left(\textbf{x}, \textbf{v}\right), \label{eq:plant_evolution}\\
 \textbf{y} &= \textbf{g}\left(\textbf{x}\right), \label{eq:plant_output}
\end{align}
where $\mathbf{x}$ are the plant internal states, $\mathbf{y}$ are the $m$ plant observables, and $\textbf{v}$ are the $l$ accessible inputs.  Generally, $\mathbf{f}$ and $\mathbf{g}$ are unknown, and the only information available is the simultaneous response of the plant to a user-defined input signal $\mathbf{v}_{train}$ during a training period.  In the following analysis, we assume that $\mathbf{f}$ is Lipschitz continuous with respect to $\mathbf{x}$, and that $\textbf{g}$ is `typical' in the sense defined by Takens' embedding theorem \cite{takens1981detecting}, so that an observer with memory of a history of $\textbf{y}(t)$ can construct a map to $\textbf{x}(t)$. There may also be noise terms included in Eq. 1 and 2, but we do not explicitly model noise in the analysis below.

Designing a controller requires an operation that reads a reference signal $\textbf{r}$ and outputs a control signal $\textbf{v}$ such that $\textbf{y}\rightarrow \textbf{r}$. It is a  \textit{closed-loop} controller if $\textbf{v}$ is also a function of the current value of $\textbf{y}$ as measured at the the plant.

If $\textbf{v}$ is constant over an interval from $t$ to $t+\delta$, where $\delta$ is a non-infinitesimal interval, then $\textbf{f}\left(\cdot,\textbf{v}\right)=\textbf{f}_\textbf{v}(\cdot)$ may be viewed as a differential equation parameterized by $\textbf{v}$. The Lipshitz condition implies that the value of $\textbf{x}\left(t+\delta\right)$ is determined by the initial conditions at $t$, \textit{i.e.},
\begin{equation}
\textbf{x}\left(t+\delta\right) = \textbf{F}_{\textbf{v}}\left[\textbf{x}\left(t\right)\right], \label{eq:evolve_x_delta}
\end{equation}
where $\mathbf{F}_\mathbf{v}$ is a nonlinear evolution operator mapping $\mathbf{x(t)}$ to $\mathbf{x(t+\delta)}$.  It can be constructed approximately by repeated application of $\mathbf{f}$ over infinitesimal time steps.

If $\textbf{v}$ is instead slowly varying from $t$ to $t+\delta$, then we expect this equality to instead be an approximation given by
\begin{equation}
\textbf{x}\left(t+\delta\right) \approx \textbf{F}\left[\textbf{x}\left(t\right), \textbf{v}\left(t\right)\right] \label{eq:evolve_x_delta_approx}
\end{equation}
for some function $\textbf{F}$. In general, this function is not invertible since there may be multiple possible input trajectories $\textbf{v}\left(t\right)$ that drive the system to a given future state and not all states may be reachable from the current state.  However, if we restrict the domain to future states reachable from the current state, we can effectively invert this function as
\begin{equation}
\textbf{v}\left(t\right) \approx \textbf{F}^{-1}\left[\textbf{x}\left(t\right), \textbf{x}\left(t+\delta\right)\right], \label{eq:v_Finverse}
\end{equation}
where $\textbf{F}^{-1}$ is the function of interest for devising a controller for Eqs.~\ref{eq:plant_evolution} and \ref{eq:plant_output} that chooses a principle value among possible inputs.  In Sec. \ref{sec:one_layer}, the future value $\textbf{x}(t + \delta)$ is replaced with a desired future value calculated at the current time $t$ in order to obtain a causal control law.

\section{\label{sec:RCintroduction}Reservoir Computing}

In RC, a recurrent network of time-dependent nodes and fixed connection weights (known as the reservoir) is driven with an input signal of interest. The response of the reservoir is then recorded, and a time-independent transformation is obtained to map the reservoir state onto a target signal. This prescription requires that the reservoir be sufficiently high-dimensional and complex so that distinct input signals are effectively separated in the state space of the reservoir. It also requires that the system exhibit the echo-state property \cite{jaeger2001} (or, more generally, generalized synchronization \cite{lu2018}) with respect to the input signal. This ensures that the reservoir exhibits short-term memory, while eventually forgetting past values of the input signal.

Although many types of physical systems with different topologies have been proposed as the reservoir substrate (see Ref. \cite{tanaka2019recent} for a thorough review), it is common to use a recurrent neural network (RNN). Within this paradigm, a reservoir computer has three distinct components: a feed-forward input  layer, a recurrent layer, and a feed-forward output layer. The input-to-reservoir connection weights $\textbf{W}_{in}$ and the reservoir-to-reservoir weights $\textbf{W}$ are randomly assigned to initial values and kept fixed. The network is then driven by an input signal $\textbf{y}(t)$ during a training period during which the response of the reservoir $\textbf{u}(t)$ is observed. Finally, given a desired output signal $\textbf{v}_d(t)$, the reservoir-to-output weights $\textbf{W}_{out}$ are chosen so that $\textbf{v}(t) \simeq \textbf{v}_d(t)$ during the training period after some initial transient is discarded.  Because training only involves linear optimization of the output layer, there is no vanishing gradient problem that makes training deep learning networks difficult.

\subsection{Echo State Networks}\label{sec:ESNintro}

The reservoir of an ESN is an $N$-node recurrent ANN whose behavior is described by the differential equation 
\begin{align}
 c\dot{\textbf{u}} &= -\textbf{u} +  \textbf{tanh}\left(\textbf{W}\textbf{u}+\textbf{W}_{in}\textbf{y}+\textbf{b} \right), \label{eq:ESN}\\
 \textbf{v} &= \textbf{W}_{out}\textbf{u}, \label{eq:ESNoutput_layer} 
\end{align}
where $\mathbf{u}$ is an $N$-dimensional vector and $\textbf{tanh}$ is the vectorized function, $\textbf{W}$ (dimension $N\times N$) is the adjacency matrix for the internal reservoir links, $\textbf{W}_{in}$ ($N \times m)$ is the input-weight matrix, and $\textbf{b}$ ($N$) is a bias vector. They are composed of random matrix elements that are fixed at the initialization of the reservoir and describe the reservoir dynamics with respect to an input signal $\textbf{y}$. The only trained parameter is $\textbf{W}_{out}$ ($l \times N$).

After driving the ESN and observing the reservoir response, we use Tikhonov regularization to determine $\textbf{W}_{out}$ by minimizing
\begin{equation}
\left| \textbf{v}_d(t) - \textbf{W}_{out}\textbf{u}(t)\right|^2 + \left|\beta\textbf{W}_{out}\right|^2,\label{eq:Tikhonov}
\end{equation}
from $t=T_{init}$ to $t=T_{train}$, where $T_{init}$ is long enough to get beyond the transient response of the reservoir, $T_{train}$ is the end of the training period, and $\beta$ is a small regularization parameter chosen to prevent overfitting to data and can be chosen with standard cross-validation techniques.

The fixed parameters $\textbf{W}, \textbf{W}_{in}, \textbf{b}$ are instantiated according to a number of hyperparameters that are selected prior to training.  These are: the magnitude of the largest eigenvalue (also known as the spectral radius) of $\textbf{W}$, $\rho$; the proportion $k$ of nonzero elements of $\textbf{W}$, which is also the mean in-degree of the network; the scale $\sigma$ of $\textbf{W}_{in}$; the mean $b_{mean}$ and maximum $b_{max}$ values of $\textbf{b}$; and the time constant $c$. They are often selected by hand based on some heuristics \cite{lukovsevivcius2012practical}, but may also be optimized by various algorithms \cite{griffith2019forecasting}. In all cases, we find that typical choices for most of these parameters provide good performance and do not explore other values in detail.

\section{Single-Layer Reservoir Controller}\label{sec:one_layer}

Direct inverse control \cite{norgaard2000neural} involves modeling the relationship in Eq.~\ref{eq:v_Finverse} with some physical assumptions about the plant and observation measurements $\{\textbf{y}(t), \textbf{v}(t); 0 \leq t \leq T \}$. The function $\textbf{F}^{-1}$ is used to devise a closed-loop controller by replacing $\textbf{x}(t+\delta)$ with the desired plant state. However, this function depends on the internal state $\textbf{x}$, whereas only the observation $\textbf{y}$ is available to the controller. In a general scenario, this signal may be missing important information about $\textbf{x}$, such as when $\textbf{g}$ projects $\textbf{x}$ onto a lower-dimensional space.

The operating principle behind ESNs is their ability to synchronize, in a generalized sense, with their inputs \cite{lu2018}. This means that a reservoir coupled to $\textbf{y}(t+\delta)$ and $\textbf{y}(t)$ will tend towards some function of $\textbf{x}(t+\delta)$ and $\textbf{x}(t)$. If we denote the reservoir state by $\textbf{u}(t)$, then for some (unknown) function $\textbf{G}$ 
\begin{equation}
\lim_{t\rightarrow \infty}\textbf{u}(t) = \textbf{G}\left[\textbf{x}\left(t+\delta\right), \textbf{x}\left(t\right)\right].
\end{equation}
Equivalently, $\textbf{u}(t)$ is approximately a function of $\textbf{x}(t+\delta)$ and $\textbf{x}(t)$ after some appropriate waiting time $T_{init}$.

Using this synchronization property for a reservoir with high enough dimension, then we can find a $\textbf{W}_{out}$ such that
\begin{equation}
\textbf{W}_{out}\textbf{G}\left[\textbf{x}\left(t+\delta\right), \textbf{x}\left(t\right)\right] \approx \textbf{F}^{-1}\left[\textbf{x}\left(t+\delta\right), \textbf{x}\left(t\right)\right] \label{eq:v_approx_Finverse}
\end{equation}
where the left hand side is the trained reservoir output that approximately `inverts' the plant dynamics.

\begin{figure}[h!]
    \centering
    \includegraphics[width=0.45\textwidth]{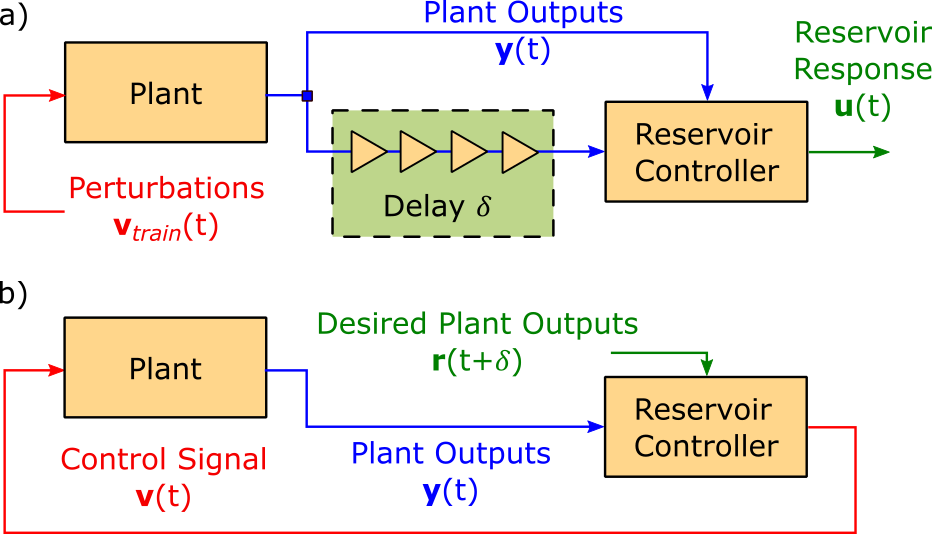}
    \caption{A schematic of the plant and reservoir controller in the a) training and b) control configurations.}
    \label{fig:control_schematic}
\end{figure}

The training data is generated by perturbing the plant with a small random signal applied to the inputs $\textbf{v}_{train}$ from $t=0$ to $t=T_{train}+\delta$, which ensures the plant is stimulated with many frequencies to explore the complete phase space. During this time, triplets $\textbf{y}(t+\delta)$, $\textbf{y}(t)$, and $\textbf{v}_{train}(t)$ are collected and used to train an ESN with $\textbf{v}_d = \textbf{v}_{train}$. The configuration of the plant and ESN in this training phase is depicted in Fig.~\ref{fig:control_schematic}a.

Through this training, the reservoir learns to approximately invert the internal plant dynamics using $\textbf{y}$ without the intermediate step of learning  $\textbf{F}^{-1}$. To control the plant, $\textbf{y}(t+\delta)$ is replaced with $\textbf{r}(t+\delta)$, where $\textbf{r}(t)$ is the desired behavior of the plant. If the ESN has learned $\textbf{F}^{-1}$, then the resulting $\textbf{v}(t)$ is precisely the control signal that drives $\textbf{y}(t+\delta) \rightarrow \textbf{r}(t+\delta)$. The complete dynamics of the controlled plant are then described by Eqs.~\ref{eq:plant_evolution} and \ref{eq:plant_output} with Eq.~\ref{eq:ESN} replaced by
\begin{equation}
 c\dot{\textbf{u}} = -\textbf{u} +  \textbf{tanh}\left(\textbf{W}\textbf{u}+\textbf{W}_{in}^y\textbf{y}+\textbf{W}_{in}^r\textbf{r}_{\delta}+\textbf{b} \right).\label{eq:control_reservoir}
\end{equation}
For simplicity and clarity, we write $\textbf{r}(t+\delta) \equiv \textbf{r}_{\delta}$ and split the input weights as $\textbf{W}_{in}^y$ and $\textbf{W}_{in}^r$, the latter of which couples to $\textbf{y}(t+\delta)$ in the training phase and $\textbf{r}(t+\delta)$ in the control phase. The configuration of the plant and ESN in this control phase is in Fig.~\ref{fig:control_schematic}b.

In physical implementations, driving the reservoir with $\textbf{y}(t)$ and $\textbf{y}(t+\delta)$ can be accomplished with a delay line as shown in Fig.~\ref{fig:control_schematic}a. This couples $\textbf{W}_{in}^y$ to $\textbf{y}(t-\delta)$ and $\textbf{W}_{in}^r$ to $\textbf{y}(t)$, which is the desired configuration under a shift $t \rightarrow t + \delta$ done after the training phase is complete.

As we discuss below, the control algorithm is capable of controlling a wide range of systems. However, $|\textbf{y}(t)-\textbf{r}(t)|$ does not converge to 0. This is to be expected because the reservoir computer only approximately learns $\mathbf{F}^{-1}$.  In the RC literature, it is known that learning improves with increasing $N$ and hence this is a strategy to reduce the control error. However, as we show in Sec.~\ref{sec:control_hyperparameters}, increasing $N$ generally decreases $|\textbf{v}(t) - \textbf{v}_{train}|$ but not $|\textbf{y}(t)-\textbf{r}(t)|$. For situations where precise control is critical, an algorithm for improving the control error is desired.  One of our contributions is to iteratively execute the control algorithm in a parallel configuration as described in Sec.~\ref{sec:control_layers}.

\subsection{Choosing $\textbf{v}_{train}$}

During the training phase, we need to specify the training signal $\textbf{v}_{train}$ (dimension $l$) when the plant and reservoir computer are in the training phase shown in Fig.~\ref{fig:control_schematic}a.  Identifying an optimal perturbation signal is an important problem in system identification and a number of methods have been developed \cite{rivera2003plant}. In keeping with the spirit of the RC framework, we randomly generate $\textbf{v}_{train}$ according to a number of control-specific hyperparameters.

Recall that Eq.~\ref{eq:evolve_x_delta_approx} holds if $\textbf{v}$ varies slowly with respect to $\delta$. This suggests that $\textbf{v}_{train}$ be bandwidth limited with frequency cutoff $1/\lambda$ with $\lambda > \delta$. Another consideration is the magnitude of perturbations $p$. Generally, large perturbations will be easier to learn because they have a greater effect on the plant. However, this may not be the best way to learn to control the plant and real-world control applications often require bounded inputs.  Our approach is to generate a training signal from a uniform random distribution, which is Fourier-transformed and frequencies above $1/\lambda$ are dropped. The signal is then inverse-Fourier-transformed, and scaled to the range $[-p,p]$ yielding $\textbf{v}_{train}$ with the required properties. 

In addition to the usual hyperparameters discussed in the previous section, we also must optimize the control-specific hyperparameters. In the next subsection, we explain how we select these hyperparameters based on the physical properties of the plant in the context of controlling the Mackey-Glass chaotic system. Additionally, we study the effect of $N$ and come to the surprising conclusion that increasing $N$ does not result in increased control performance.

\subsection{Hyperparameter Considerations: The Mackey-Glass System}\label{sec:control_hyperparameters}

The Mackey-Glass system is described by a delay-differential equation, which is augmented by an additive drive signal $v(t)$ and an observer as 
\begin{align}
\dot{x}(t) &= - \gamma x(t) + \beta_{MG} \frac{x(t-\tau)}{1+x^q(t-\tau)} + v(t), \label{eq:MG}\\
y(t) &= x(t). \label{eq:MG_observer}
\end{align}
We take $\beta_{MG}=0.2$, $\gamma=0.1$, $q=10$, and $\tau=17$, which places Eq.~\ref{eq:MG} in the chaotic regime without input.  We simulate Eq. 11-13 with a 4$^{th}$-order Runge-Kutta method and fixed step size $h=0.1$.

We investigate the effect of hyperparameter selection by attempting to stabilize the unstable steady state (USS) $x(t) = 1$ of Eq.~\ref{eq:MG}. That is, after $\textbf{W}_{out}$ is identified, the control configuration is obtained by replacing and $\textbf{r}_\delta$ in Eq.~\ref{eq:control_reservoir} with the constant $1.0$ and $v(t)$ in Eq.~\ref{eq:MG} with the trained reservoir output. Motivated by the training algorithm to find $\mathbf{W}_{out}$, the first quantity of interest is the normalized difference between the training and actual input signals after training has been completed and during the interval $[T_{train},T_{test}]$ given by 
\begin{equation}
     \sqrt{ \frac{1}{T_{test}\text{var}(v)} \int_{t=T_{train}}^{T_{train}+T_{test}}\left( v(t) - v_{train}(t) \right)^2dt}.
\end{equation}
Because $v$ is approximately equal to $F^{-1}$ (recall Eq.~\ref{eq:v_approx_Finverse}), this metric is approximately equal to the error in finding the inverse plant dynamics.  For brevity, we refer to this simply as the plant inversion error, although it is understood that this is only an approximation. 

We also measure the control error once control-loop feedback is enabled during the control configuration shown in Fig.~\ref{fig:control_schematic}b.  The asymptotic control error is given by
\begin{equation}
  \lim_{T \rightarrow \infty}
   \frac{1}{T} \int_{T_{control}}^{T_{control}+T}
  \left|y(t) - r(t)\right|dt. \label{eq:control_error}   
\end{equation}
There is a possibility that the plant inversion error and the control error are different because the former is in an open-loop configuration (Fig.~\ref{fig:control_schematic}a) whereas the later is in a closed-loop configuration (Fig.~\ref{fig:control_schematic}b).  

\begin{figure}[h]
    \centering
    \includegraphics[width=0.5\textwidth]{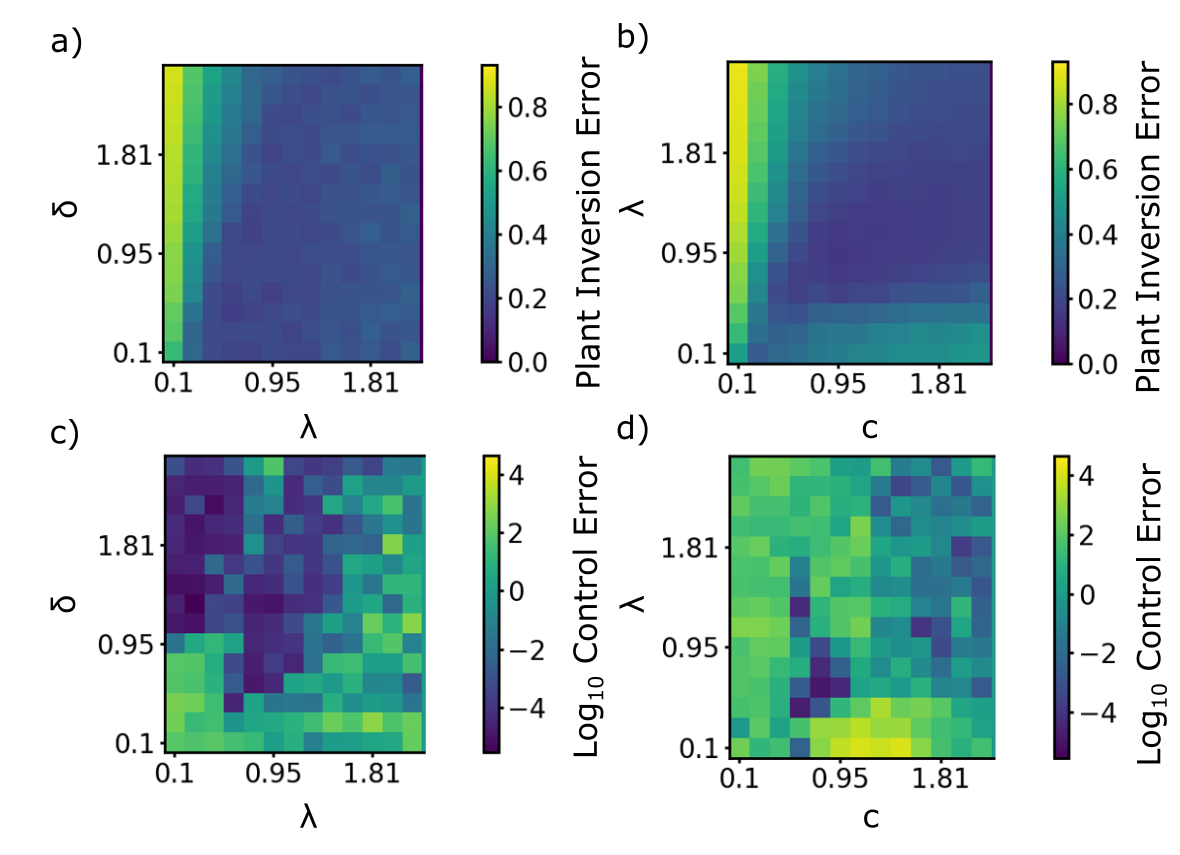}
    \caption{Parametric dependence of the RC-based controller a) and b) plant inversion error with $c$=0.6, and c) and d) control error with $\delta$=0.6 when stabilizing an USS of the Mackey-Glass system.}  
    \label{fig:tune_control_parametres}
\end{figure}

Unless otherwise specified, the RC and control hyperparameters are given in Table 1. As discussed in the previous subsection, the range of $p$ is often restricted by case-specific constraints. The parameters $\delta$ and $\lambda$ are particularly interesting in that they introduce two additional temporal scales, where the typical RC problem only contains $c$.  Above, we argue that $\lambda > \delta$ is expected for good plant inversion error. Similarly, we expect that $\lambda \approx c$ because the reservoir nodes themselves are frequency filters with cut-off $1/c$. We test these ideas by simultaneously varying the temporal parameters as shown in Fig.~\ref{fig:tune_control_parametres}.  

\begingroup
\renewcommand{\arraystretch}{1.5}
\begin{table}[]
    \centering
    \begin{tabular}{|c|c|c|c|c|c|c|}
      \hline
      Parameter & $N$ & $\rho$ & $k$ & $\sigma$  & $b_{max}$ & c \\
      \hline
      Value & 100 & 1.15 & 10 & 1 & 1 & 0.6 \\ 
      \hline 
      \hline
      Parameter & $\delta$ & $\lambda$ & $p$ & $T_{init} $ & $T_{train}$ & $\beta$ \\
      \hline
      Value & 0.6 & 0.6 & 0.1 & 100 & 1,500 & 10$^{-8}$ \\
      \hline
    \end{tabular}
    \vspace{10pt}
    \caption{The hyperparameters used to control the Mackey-Glass system, unless otherwise noted.}
    \label{tab:MGparameters}
\end{table}
\endgroup

We adjust the parameters while choosing $\mathbf{W_{out}}$ to minimize the inversion error. As seen in Fig.~\ref{fig:tune_control_parametres}a and b, the plant inversion error is a relatively smooth function in this parameter space, with minima in the $(\lambda, \delta)$ plane below the $\lambda = \delta$ line and minima in the $(\lambda, c)$ plane along the $\lambda = c$ line.  The false-color plots are on a linear scale.

On the other hand, Fig.~\ref{fig:tune_control_parametres}c and d reveal that the  control error has a more complex dependence on the parameters and that different parameter combinations result in low error. Also, the observed variation in the control error is larger than the inversion error - the false-color scale is on a logarithmic scale - making it difficult to arrive at a simple functional dependence of the error landscape on the parameters $\delta$, $\lambda$, and $c$.

\begin{figure}[h]
    \centering
    \includegraphics[width=0.3\textwidth]{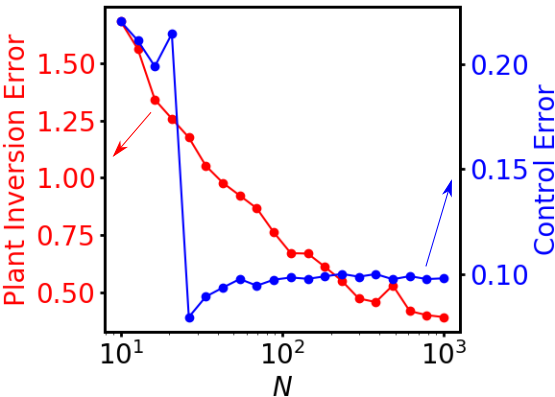}
    \caption{Performance of a single-layer reservoir-computer controller when applied to the Mackey-Glass equations.} 
    \label{fig:error_vs_N}
\end{figure}

Finally, we investigate the effect of reservoir size $N$.  Based on previous RC computing research, we expect that the approximate plant inversion will decrease with $N$ because the training algorithm (Eq.~\ref{eq:Tikhonov}) is designed to choose $\textbf{W}_{out}$ to minimize this error.  This is confirmed in Fig.~\ref{fig:error_vs_N}. 

We find a very different dependence on the control error.  There is a sudden drop in the error with increasing $N$ corresponding to a bifurcation of the closed-loop feedback + plant system. For $N$ too small, the attractor of the controlled Mackey-Glass system is identical to the uncontrolled Mackey-Glass system. However, there is a sudden change at a critical $N$, beyond which the attractor of the controlled system becomes a fixed point very close to the requested USS. We see similar behaviors when controlling the other systems described later.  We also observe a minimum control error near $N=30$ immediately after the bifurcation.  Thus, there is little benefit to controlling $N$ beyond this value and there is even a slight increase in the error for larger $N$. 

Our findings also point to the challenge in controlling high-complexity systems that display chaos.  A single reservoir-computer controller has been shown previously to be effective in controlling the behavior of heated and stirred tank, pitch control of an aircraft, a double pendulum \cite{waegeman2012feedback}, and a robot arm \cite{waegeman2012feedback}.  While it might be possible to improve the controller performance studied here by fine-tuning the reservoir and control hyperparameters, we introduce an alternative method in the next section. 

\section{The Deep ESN Controller}\label{sec:control_layers}

In this section, we introduce a layered approach to obtain lower control error. As motivation, consider Eq.~\ref{eq:control_reservoir} for the controlled plant.  It can be thought of as another (partially) unknown dynamical system with internal state given by $\{\textbf{x, u}\}$ and output $\textbf{y}$. An accessible control input $\textbf{v}'$ can be created in a number of ways, such as with the replacement $\textbf{v} \rightarrow \textbf{v} + \textbf{v}'$. Because this new plant is partially controlled, the trajectory of $\textbf{y}$ is now much closer to $\textbf{r}$ than in the uncontrolled system given by Eqs.~\ref{eq:plant_evolution} and \ref{eq:plant_output}. This means that the partially controlled system is generally easier to control with the same strategy described above.

\begin{figure}[h]
    \centering
    \includegraphics[width=0.35\textwidth]{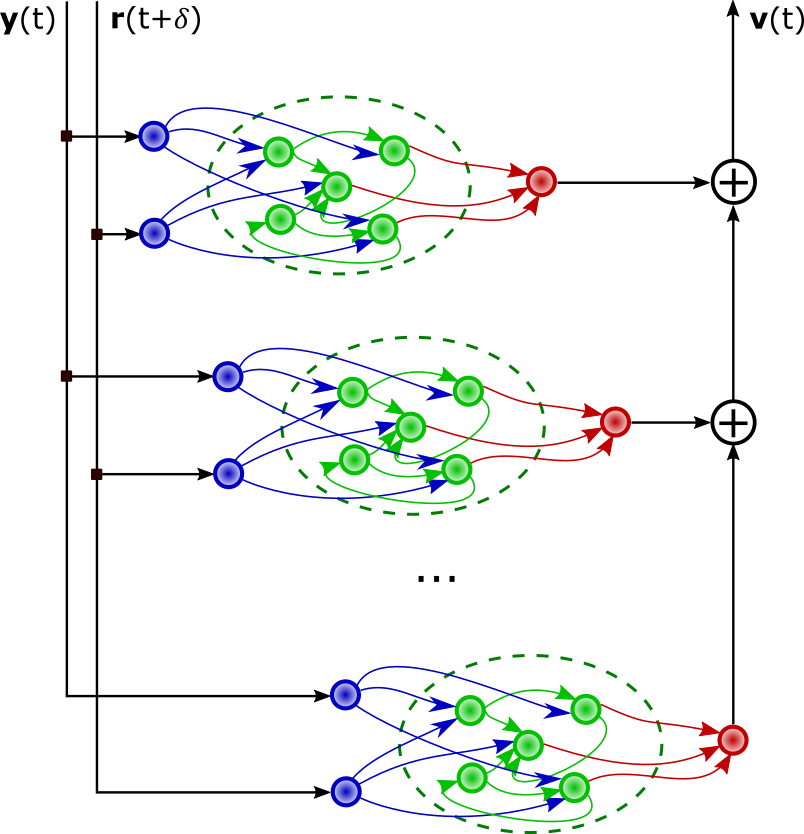}
    \caption{The configuration of the $n_{th}$ reservoir controller. All layers of the controller take as input $\textbf{y}$ and $\textbf{r}_\delta$, which couple to the $i_{th}$ reservoir through $\textbf{W}_{in,i}^y$ and $\textbf{W}_{in,i}^r$, respectfully. The trained weights $\textbf{W}_{out,i}$ depend only on the measured dynamics of the $(i-1)_{th}$ controller, so the deep controller is trained sequentially. }
    \label{fig:dESN}
\end{figure}

The layered controller is described by Eqs.~\ref{eq:plant_evolution} and \ref{eq:plant_output}, and 
\begin{align}
    c_i \dot{\textbf{u}}_i &= -\textbf{u}_i + \textbf{tanh}\left(\textbf{W}_i\textbf{u}_i+\textbf{W}_{in, i}^y\textbf{y}+\textbf{W}_{in, i}^r\textbf{r}_{\delta}+\textbf{b}_i \right), \\ \nonumber
    &\text{~~~~~~~~~~~~~~~~~~for } 1 \leq i \leq n,  \\
    \textbf{v} &= \sum_{i=1}^{n}
    \textbf{W}_{out, i}\textbf{u}_i, 
\end{align}
where we simply add the outputs from each reservoir and $\textbf{W}_{out, i}$ is trained by controlling the $\left(i-1\right)_{th}$ controlled plant as described below.  The deep controller is illustrated in Fig.~\ref{fig:dESN}.

Adding additional control layers adds additional hyperparameters to consider.  In general, the hyperparameters should be selected for each layer, but we find that low control error is obtained using the same parameters for all layers, thereby greatly simplifying the design process. We follow this approach in the additional examples given below. 

An added benefit of the layered approach is its computational efficiency.  The training algorithm given in Eq.~\ref{eq:Tikhonov} scales approximately as $N^3$ in the size of the reservoir, but only linearly in the number of reservoirs.  Thus, it is more efficient to train several smaller independent reservoir computers than a single large reservoir computer with the same number of total nodes.

\section{Controlling the Lorenz System}\label{sec:LorenzResults}

In this section, we use our algorithm to control the multi-input multi-output Lorenz '63 system, described by
\begin{align}
\dot{x_1} &= \sigma_L \left(x_1 - x_2\right) + v_1, \label{eq:Lorenzx1}\\
\dot{x_2} &= x_1\left(\rho_L - x_3\right) - x_2 + v_2, \label{eq:Lorenzx2} \\
\dot{x_3} &= x_2x_2 - \beta_L x_3 + v_3, \label{eq:Lorenzx3}\\
\textbf{y} &= \textbf{x}, \label{eq:Lorenzy}
\end{align}
where $\mathbf{x}=(x_1,x_2,x_3)$ and $\mathbf{v}=(v_1,,v_2,v_3)$. We consider the typical parameters $\sigma_L=10$, $\rho_L=28,$ and $\beta_L=8/3$, for which Eqs.~\ref{eq:Lorenzx1}-\ref{eq:Lorenzy} display chaotic behavior when $\mathbf{v}=\textbf{0}$.  We solve numerically the Lorentz and RC-based controller equations using a 4$^{th}$-order Runge-Kutta method with fixed step size of 10$^{-3}$. 

For $\mathbf{u}=\textbf{0}$, unstable steady states exist at $\left(0,0,0\right)$ and $\left(\rho_L, \pm \sqrt{\rho_L \beta_L}, \pm \sqrt{\rho_L \beta_L}\right)$, the latter of which exist at the center of the symmetric leaves of the attractor.  We find that the dESN controller can stabilize or induce a wide variety of behaviors and give only two examples here for brevity.

\subsection{Controlling an USS of the Lorenz System Using a Single-Layer Controller} \label{sec:LorenzUSS}

We first use the dESN controller to stabilize the Lorenz system to the positive USS using a single control layer with the parameters in Table 2.  Because the USS is an unstable set of the attractor, stabilizing the attractor to the USS should require control perturbations whose size are dictated by the noise level, which is set only by the numerical integration error in our simulations and much smaller than the errors observed below.

\begingroup
\renewcommand{\arraystretch}{1.5}
\begin{table}[h]
    \centering
    \begin{tabular}{|c|c|c|c|c|c|c|c|}
      \hline
      Parameter & $N$ & $\rho$ & $k$ & $\sigma$  & $b_{max}$ & c \\
      \hline
      Value & 200 & 0.9 & 20 & 0.05 & 1 & 0.01 \\ 
      \hline 
      \hline
      Parameter & $\delta$ & $\lambda$ & $p$ & $T_{init} $ & $T_{train}$ & $\beta$ \\
      \hline
      Value & 0.05 & 0.05 & 10 & 25 & 250 & 10$^{-8}$ \\
      \hline
    \end{tabular}
    \vspace{10pt}
    \caption{The hyperparameters used to control the Lorenz system to the positive USS, unless otherwise specified.}
    \label{tab:Lorenz_parameters}
\end{table}
\endgroup

\begin{figure}[h!]
    \centering
    \includegraphics[width=0.48\textwidth]{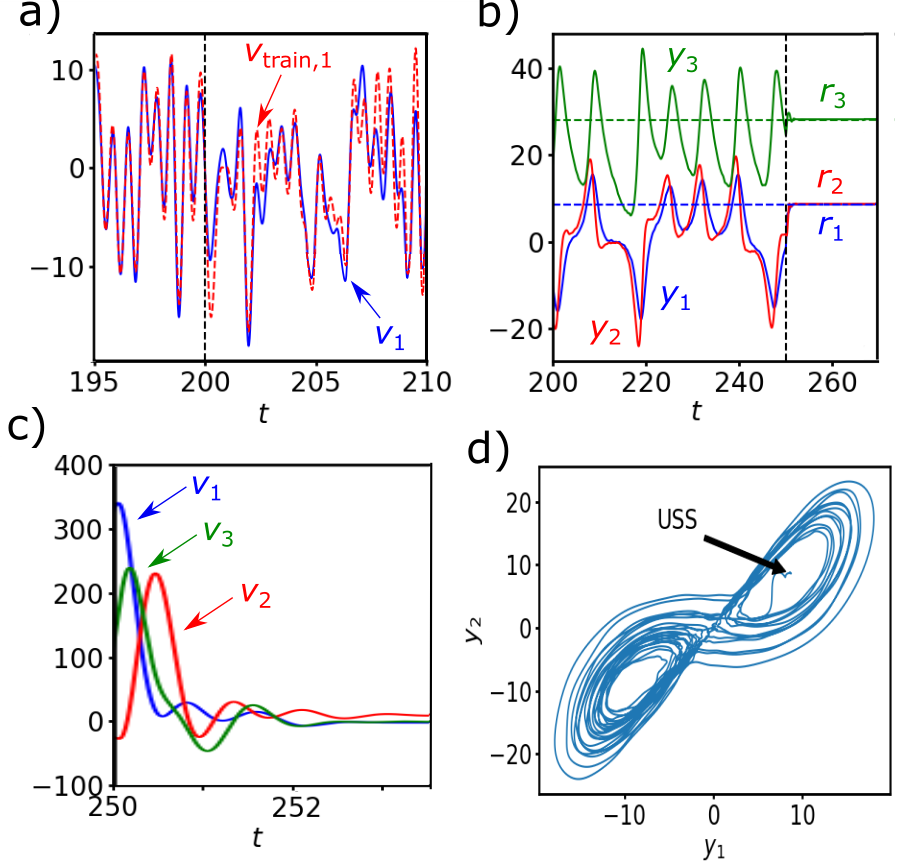}
    \caption{Control of the Lorenz system to the positive USS. a) The first component $v_1$ of the reservoir output/control input compared to the first component $v_{train,1}$ of the training input to Lorenz. b) Dynamics of the Lorenz system before and after the controller is switched on at $t=250$. c) The control signal, as generated by the trained reservoir. d) The Lorenz system in phase space.}
    \label{fig:LorenzUSS}
\end{figure}

The plant and reservoir dynamics for a single-layer controller are shown in Fig.~\ref{fig:LorenzUSS}.  During the training phase, the reservoir computer does a reasonable job of inverting the plant dynamics as seen in Fig.~\ref{fig:LorenzUSS}a.  Here, we show how well the reservoir computer inverts the plant when we replay the training data ($t<200$) through the system after the weights have been trained and its ability to generalize to a noisy drive signal it has not seen previously ($t>200$).  The normalized root-mean-square error during the training time is 0.50.  

When control is turned on at $t=250$ in Fig.~\ref{fig:LorenzUSS}b, it rapidly approaches the desired USS, which can be seen more easily in the two-dimensional projection of the attractor in phase space shown in Fig.~\ref{fig:LorenzUSS}d.  The control perturbations are quite large initially (Fig.~\ref{fig:LorenzUSS}c) - comparable to the size of the other terms on the right-hand side of Eqs.~\ref{eq:Lorenzx1}-\ref{eq:Lorenzx3} in the absence of control - but then decrease rapidly.  We make no attempt to minimize the initial size of the control perturbations; in the case shown, the controller essentially stops the chaotic dynamics in its tracks and then guides the trajectory to the desired USS.

Standard chaos control methods that require only small perturbations \cite{Shinbrot1993} cannot be applied to control this USS because they require that the chaotic trajectory passes within a neighborhood of the desired unstable state at which point control is turned on.  This demonstrates the ability of our control approach to stabilize behaviors that are not on the attractor, albeit at the cost of larger control perturbations. 

A close inspection of Fig.~\ref{fig:LorenzUSS}c reveals that the control perturbations do not go to zero as expected.  This is due to the small error in the reservoir computer estimating the plant inverse.  As with controlling the Mackey-Glass system (Sec.~\ref{sec:one_layer}), we find that increasing $N$ does not improve the control performance even though it decreases the plant inversion error (data not shown).  We thus add layers to the controller to improve its performance.

\subsection{Applying the dESN to the Lorenz System}

We control the Lorenz attractor to the same USS discussed in the previous section using progressively more layers with the same hyperparameters as in Table~\ref{tab:Lorenz_parameters} except that each layer has $N=50$. The state of the Lorenz system and the real-time control error from a typical example are presented in Fig. 6. Here, we include both the training and control phases as 4 layers are added to the controller.

\begin{figure*}[h]
    \centering
    \includegraphics[width=0.7\textwidth]{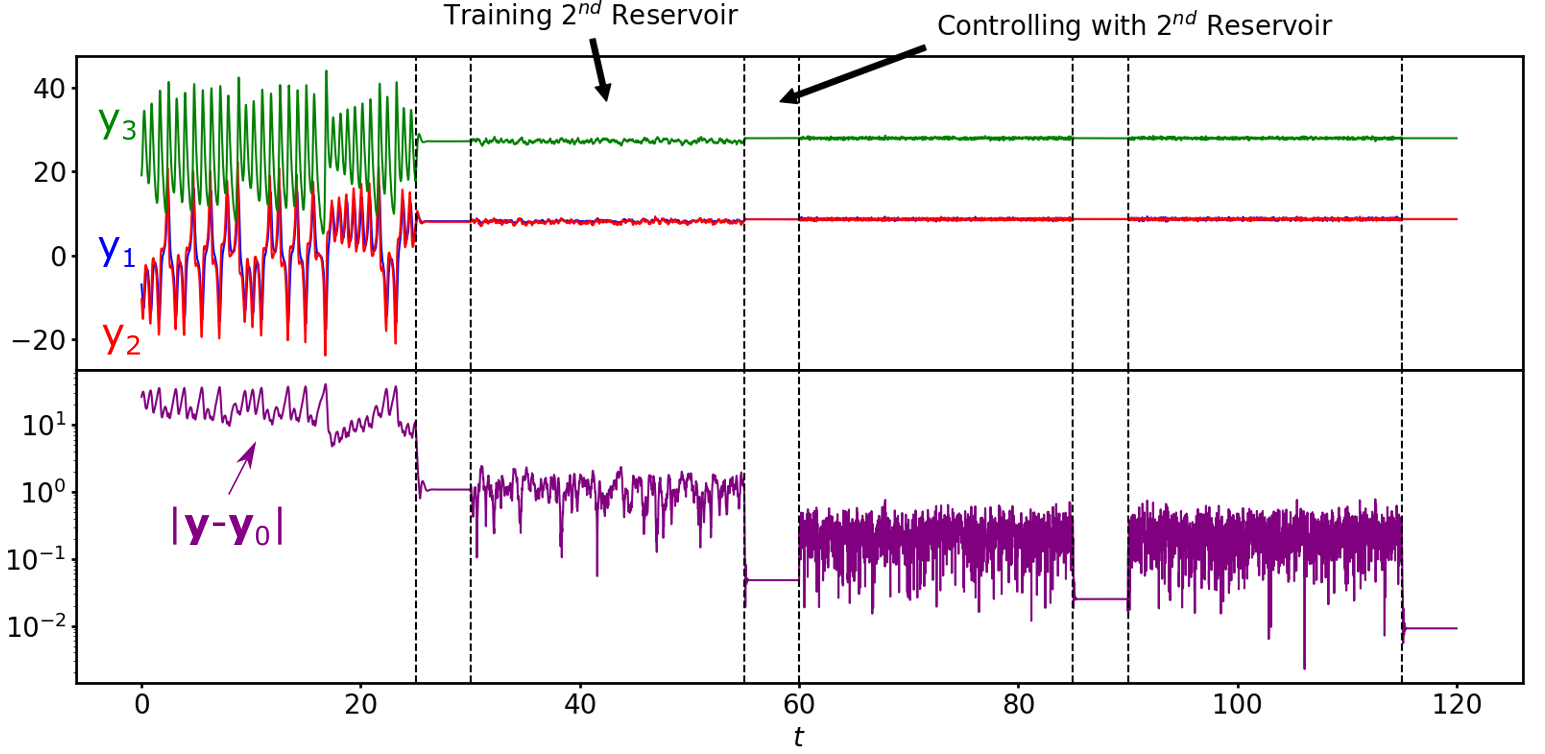}
    \caption{A typical trajectory of a controlled Lorenz system. Vertical dashed lines separate successive training and control phases, with the error from the requested USS $\textbf{y}_0$ shown in the bottom panel. That is, the system is perturbed with the first training signal from $t=0$ to $t=25$, and the single-layer controller is applied from $t=25$ to $t=30$. The controlled system is then perturbed with a second training signal from $t=30$ to $t=55$, and so on.} 
    \label{fig:Lorenz_layered_control}
\end{figure*}

As seen in the figure, each additional layer provides more precise control over the Lorenz system. After four layers (200 total nodes), the final control error is improved by two orders-of-magnitude relative to the error from the first layer.  Other simulations (data not shown) predict that a single-layer reservoir computer controller with $N=200$ does no better than a single-layer with $N=50$.  Figure~\ref{fig:Lorenz_layered_control} also demonstrates that the controlled system is highly stable to the training perturbations as each layer is added.

\subsection{Controlling Ellipses Near the Lorenz Attractor} \label{sec:Lorenz_ellipse}

\begin{figure}[h!]
    \centering
    \includegraphics[width=0.45\textwidth]{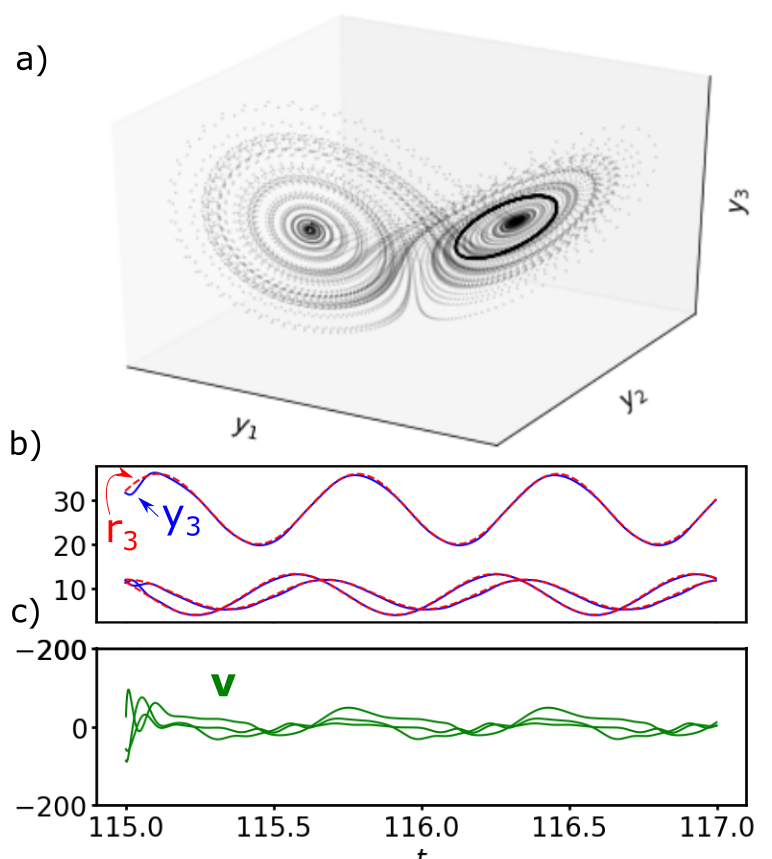}
    \caption{Controlling the Lorenz chaotic system to an elliptical orbit. a) The phase space portrait of the Lorenz system (dashed) and the requested ellipse (solid). b) The desired orbit (dashed) and the controlled Lorenz system (solid) in real space as a 4-layer controller is switched on. Note that the difference is vanishingly small except for an initial transient. c) The control perturbation of the 4-layer controller.}
    \label{fig:Lorenz_ellipse}
\end{figure}

To highlight the fully nonlinear control characteristics of the dESN, we stabilize an elliptical periodic attractor that is in the vicinity of one of the leaves of the Lorenz attractor.  It is located near the positive lobe of the attractor and centered around the positive USS, as illustrated in Fig.~\ref{fig:Lorenz_ellipse}a. The ellipse coefficients are chosen by observing a segment of the Lorenz trajectory that spends several cycles orbiting about the positive USS and fitting to an ellipse in the least-squares sense.  No elliptical periodic behavior is a solution to the autonomous Lorenz system (Eqs.~\ref{eq:Lorenzx1}-\ref{eq:Lorenzx3}), which implies that a non-vanishing controller effort is required to stabilize this behavior.

A typical control example is presented in Figure~\ref{fig:Lorenz_ellipse}b and c, which shows the temporal evolution of the control error and perturbations, respectively, needed to stabilize the elliptical periodic behavior when control is turned on at $t=115$. The parameters are the same as in the previous section on stabilizing the USS (including $N=50$ for each layer). The asymptotic control error is only 0.13. Compare this to the mean square-root of the variance of the Lorenz attractor of approximately $25.08$. As in the previous section, the control perturbations are large initially because we make no attempt to wait until the trajectory of the uncontrolled system is close to the desired ellipse.

\section{Applying the dESN Controller to a Chaotic Electronic Circuit}

\begin{figure}[h!]
    \centering
    \includegraphics[width=0.35\textwidth]{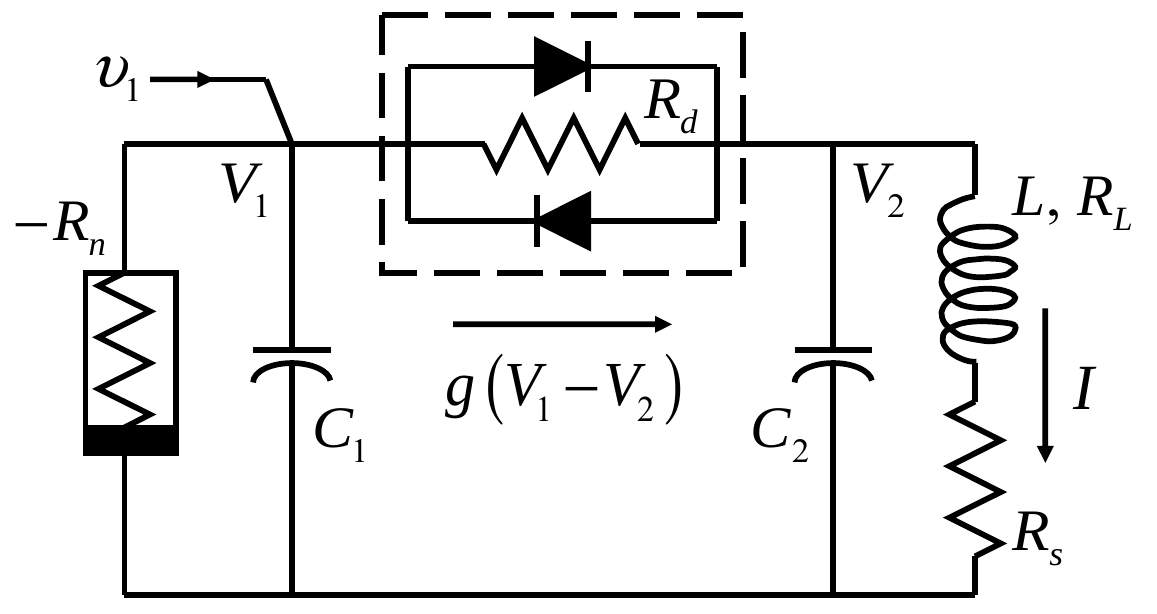}
    \caption{The chaotic electronic circuit displaying double-scroll behavior. Circuit parameters: $R_n$ = 3 k$\Omega$, $C_1$ = $C_2$ = 10 nF, $L$ = 55 mH, $R_L$ = 355 $\Omega$, $R_s$ = 100 $\Omega$, $R_m$ = $R_L$+$R_s$ = 455 $\Omega$, $R_d$ = 7.86 k$\Omega$, signal diode are type 1N914 with $V_d$ = 0.58 V with $I_r$  =5.63 nA and $\alpha$ = 11.6.  circuit to be controlled. The characteristic resistance of the circuit is $R$ = $\sqrt{R/L}$ = 2,245 $\Omega$.} 
    \label{fig:circuit}
\end{figure}

In this section, we apply the dESN controller to a chaotic electronic circuit, which consists of passive linear components, nonlinear signal diodes, and an active negative resistor \cite{chang1998stabilizing} shown schematically in Fig.~\ref{fig:circuit}. Its dynamics are governed by 
\begin{equation}
\begin{split}
    & C_1 \dot{V}_1 = \frac{V_1}{R_n} - g(V_1 - V_2) + v_1, \label{eq:circuit}\\
    & C_2 \dot{V}_2 = g(V_1 - V_2) - I + v_2, \\
    & L \dot{I} = V_2 - R_m I + v_3, \\
    & g(V) = \frac{V}{R_d} + 2I_r{\rm sinh}(\alpha \frac{V}{V_d}),
\end{split}
\end{equation}
where $V_1 (V_2)$ is the voltage drop across capacitor $C_1 (C_2)$, $I$ is the current through the inductor, $v_1 (v_2)$ is an accessible current into the $V_1 (V_2)$-node, and $v_3$ is an accessible voltage across the inductor. In the absence of control perturbations, the circuit displays double-scroll behavior for a range of negative resistances.  Similar to the Lorenz attractor discussed in Sec.~\ref{sec:LorenzResults}, the circuit has a USS at the origin and two symmetric USSs at $(\pm V_1^{ss}, \pm V_2^{ss}, I^{ss})$, with approximate values $V_1^{ss}$=0.59 V, $V_2^{ss}$=0.09 V, $I^{ss}$=0.20 mA. The error level (both electronic noise and discretization error in the analog-to-digital converter) is determined by adjusting $R_n$ so that one of the non-zero USSs is stable and measuring the root-mean-square error (RMSE) of the signal. This noise level is used in this section to contextualize the control errors.

The system described by Eq.~\ref{eq:circuit} exhibits chaotic oscillations  with a characteristic time scale of $\sqrt{L/C}$ = 23.5 $\mu$s, which is the approximate orbit time of a trajectory around the USS in the center of a scroll, and has substantial spectral components beyond 20 kHz.  Such fast time scales require a control loop with a total latency between measurement and application of the control perturbation much less than the characteristic time, which is a challenging task with a reservoir computer in the loop. To connect to the previous sections, $\textbf{x} = \left(V_1, V_2, I\right)$

To perform the measurement of the accessible system variable, evaluate the reservoir computer, and apply the control perturbations, we use a field-programmable gate array (FPGA). Specifically, we use a Max 10 10M50DAF484C6G Device on a Terasic Max 10 Plus development board. The device includes integrated dual 12-bit analog-to-digital converters for measurement that operate up to 1 MHz and a 16-bit digital-to-analog converter that operates at 1 MHz for applying the control perturbations. 

We use the ADCs to make simultaneous measurements of $V_1$ and $V_2$ as the accessible plant variables $\mathbf{y}=(V_1,V_2,0)$.  The DAC generates a voltage ($\textbf{v}_{train}$ during the training period or $\textbf{v}$ during the control period) that we send through a voltage-to-current converter with variable gain and inject into the $V_1$ node so that $\mathbf{u}=(v_1,0,0)$. 



To accelerate the dESN controller on the FPGA, we use 32-bit, fixed-point calculations and an Euler integration method for Eq.~\ref{eq:control_reservoir}. To greatly reduce hardware space, the matrices $\textbf{W}, \textbf{W}_{in}$, and $\textbf{b}$ and the time constant $c$ are hard-coded at the time of design compilation. Conversely, $\textbf{W}_{out}$ is stored in on-board RAM and updated mid-operation by a host computer. The output is then calculated by evaluating $\textbf{W}_{out} \textbf{x}$ with dedicated multipliers and adders. The tanh function is implemented with a 10-bit lookup table. The ADC, DAC, and ESNs are synchronized to a 1 MHz global clock.

\subsection{Nonlinear Control of the Chaotic Electronic Circuit}

As with the simulation results in the previous sections, we find that the dESN controller can stabilize a variety of behaviors, but focus here on the nonlinear control problem of rapidly and repeatedly guiding the system between each nonzero USS $(\pm V_1^{ss}, \pm V_2^{ss}, I^{ss})$.  The hyperparameters are selected according to the reasoning outlined in Sec.~\ref{sec:control_hyperparameters} and are listed in Tab.~\ref{tab:circuit_parameters}. 
\begingroup
\renewcommand{\arraystretch}{1.5}
\begin{table}[h]
    \centering
    \begin{tabular}{|c|c|c|c|c|c|c|c|}
      \hline
      Parameter & $N$ & $\rho$ & $k$ & $\sigma$  & $b_{max}$ & c \\
      \hline
      Value & 30 & 0.9 & 3 & 0.95 & 0.5 & 24 $\mu$s \\ 
      \hline 
      \hline
      Parameter & $\delta$ & $\lambda$ & $p$ & $T_{init} $ & $T_{train}$ & $\beta$ \\
      \hline
      Value & 8 $\mu$s & 24 $\mu$s & 22.5 $\mu$A & 512 $\mu$s & 8,192 $\mu$s & 10$^{-8}$ \\
      \hline
    \end{tabular}
    \vspace{10pt}
    \caption{The hyperparameters used to control the electronic circuit.}
    \label{tab:circuit_parameters}
\end{table}
\endgroup

\begin{figure*}[h!]
    \centering
    \includegraphics[width=0.8\textwidth]{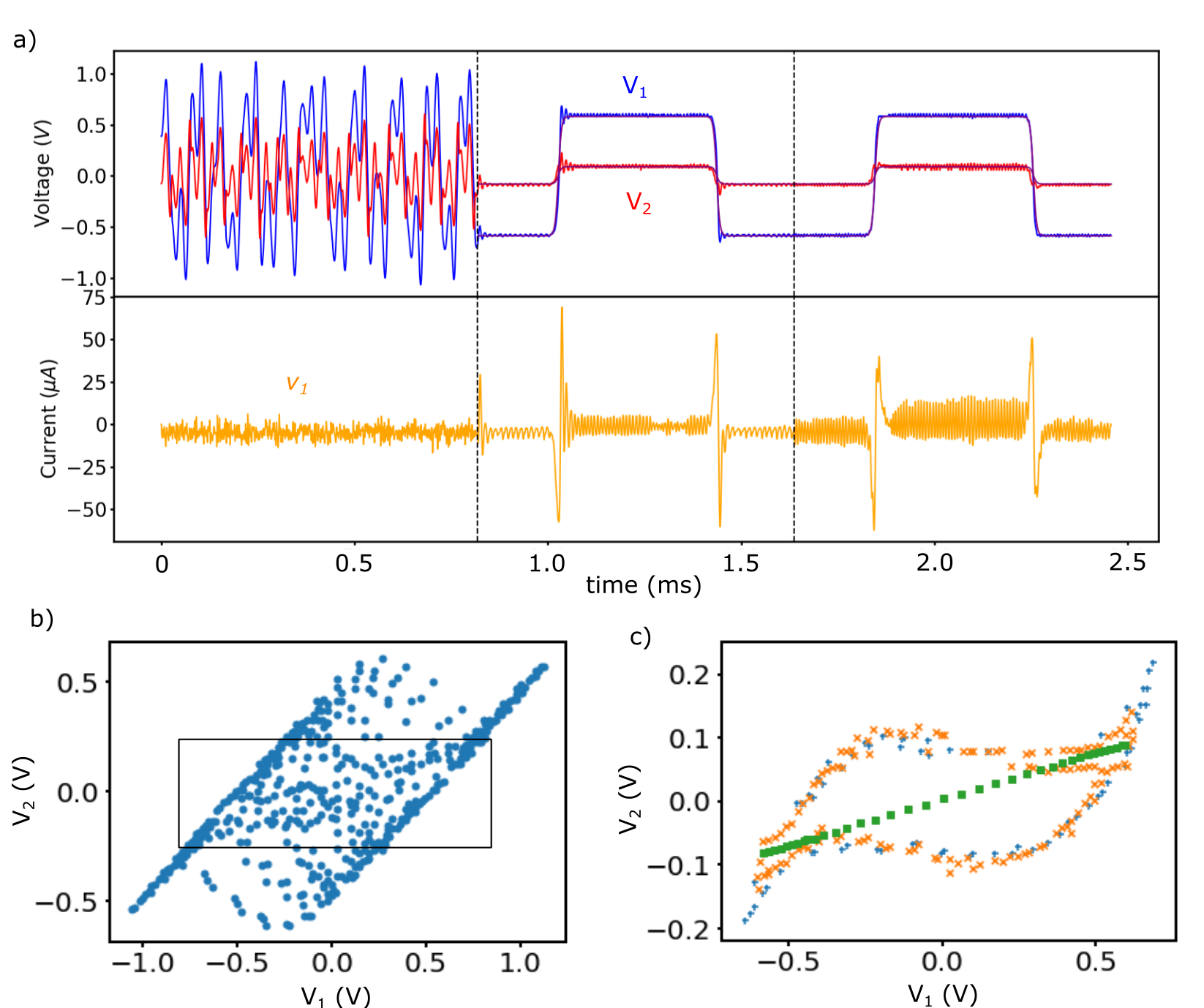}
    \caption{Control of the experimental circuit between USSs. a) In real space, the first controller leads to substantial ringing after the transition betweem USSs.  The second reservoir substantially reduces this. b) The uncontrolled attractor in phase space. The black box indicates the region containing the target trajectory in the next panel. c) In phase space, it appears that dragging straight across the attractor is an unnatural trajectory for the circuit. The circuit subject to the first controller, the circuit subject to the second controller, and the target trajectory are depicted with +, x, and square symbols, respectively.}
    \label{fig:circuit_control_betweenUSS}
\end{figure*}

Figure ~\ref{fig:circuit_control_betweenUSS}a and b show the real-space trajectories of the controlled circuit dynamics and control error, respectively, and Fig.~\ref{fig:circuit_control_betweenUSS}b shows the circuit dynamics in phase space over the same interval. As seen in Fig.~\ref{fig:circuit_control_betweenUSS}b, substantial control perturbations are required to initially control $(-V_1^{ss},-V_2^{ss},-I^{ss})$ as well as when making the transition to $(V_1^{ss},V_2^{ss},I^{ss})$ and back again.  We also observe DC offset of the system from its USSs and a ringing effect after the transition.  We also observe that requested path straight across the attractor is difficult to control precisely as can be seen most apparently in Fig.~~\ref{fig:circuit_control_betweenUSS}b.  Curiously, it appears that the DC error is largely addressed by the second reservoir. The control effort during the transition is reduced substantially by the second reservoir, but a persistent high-frequency instability is somewhat larger. 

To quantify these results, the control task is repeated a total of 30 times each with 5 different realizations of ESNs. The mean performance is characterized by the RMSE of the control error over one period and is $131.1\pm 6.9$ mV for layer 1 and $26.9\pm 0.8$ mV for layer 2. Compare these numbers to the estimated RMSE of the noise level at 13.1 mV.

\section{Conclusions}\label{sec:conclusions}

In this paper, we introduce a method for controlling an arbitrary dynamical systems to arbitrary trajectories. It requires no knowledge of the plant, and is therefore model-free. Unlike other model-free techniques, the control law is learned directly rather than through an initial system identification step. The algorithm is capable of controlling complex chaotic systems and is robust to the noise and non-ideal properties of physical systems. It can be implemented with a compact FPGA and used to control fast experimental systems. This work paves the way for research into control engineering with reservoir computing and application to real-world problems, as we have demonstrated.  The fast training time suggests that our approach can be applied to systems in real time in response to changing conditions, such as a damage event, or in manufacturing variation of system.

This research suggests several future directions in control engineering and RC more generally. First, a rigorous stability analysis is desired. While this is notoriously difficult when recurrent neural networks are involved, many safety standards require such a proof before deploying a control system when humans are involved. Second, the application of optimization methods is not well understood in this domain of RC. The issue is particularly salient here, given the increased number of hyperpameters. Particularly interesting is whether optimizations can be made by relaxing the constraint that all ESNs have the same set of hyperparameters. It may instead by the case that, say, deeper ESNs require different time constants because the local Lyapunov spectrum is different from the controlled and uncontrolled plants. Third, on the other hand, a controller with an exceptionally small footprint may be devised by considering individual reservoirs that have identical weights. Although multiple reservoirs are required during training (because the trained reservoirs couple to $\textbf{r}(t+\delta)$, while the to-be-trained reservoir couples to $\textbf{y}(t+\delta)$), they could all be replaced with a single reservoir during the control phase, where they all have the same dynamics. Initial study of this idea to the Lorenz system suggests this is possible, but more investigation is required.

\bibliographystyle{IEEEtran}
\bibliography{IEEEexample}

\begin{thebibliography}{10}
\providecommand{\url}[1]{#1}
\csname url@samestyle\endcsname
\providecommand{\newblock}{\relax}
\providecommand{\bibinfo}[2]{#2}
\providecommand{\BIBentrySTDinterwordspacing}{\spaceskip=0pt\relax}
\providecommand{\BIBentryALTinterwordstretchfactor}{4}
\providecommand{\BIBentryALTinterwordspacing}{\spaceskip=\fontdimen2\font plus
\BIBentryALTinterwordstretchfactor\fontdimen3\font minus
  \fontdimen4\font\relax}
\providecommand{\BIBforeignlanguage}[2]{{%
\expandafter\ifx\csname l@#1\endcsname\relax
\typeout{** WARNING: IEEEtran.bst: No hyphenation pattern has been}%
\typeout{** loaded for the language `#1'. Using the pattern for}%
\typeout{** the default language instead.}%
\else
\language=\csname l@#1\endcsname
\fi
#2}}
\providecommand{\BIBdecl}{\relax}
\BIBdecl

\bibitem{marino2011nested}
R.~Marino, S.~Scalzi, and M.~Netto, ``Nested pid steering control for lane
  keeping in autonomous vehicles,'' \emph{Control Engineering Practice},
  vol.~19, no.~12, pp. 1459--1467, 2011.

\bibitem{chu2018active}
Z.~Chu, Y.~Sun, C.~Wu, and N.~Sepehri, ``Active disturbance rejection control
  applied to automated steering for lane keeping in autonomous vehicles,''
  \emph{Control Engineering Practice}, vol.~74, pp. 13--21, 2018.

\bibitem{fang2011robust}
H.~Fang, L.~Dou, J.~Chen, R.~Lenain, B.~Thuilot, and P.~Martinet, ``Robust
  anti-sliding control of autonomous vehicles in presence of lateral
  disturbances,'' \emph{Control Engineering Practice}, vol.~19, no.~5, pp.
  468--478, 2011.

\bibitem{stevens2016}
B.~L. Stevens, F.~L. Lewis, and E.~N. Johnson, \emph{Aircraft control and
  simulation, 3rd Ed.}\hskip 1em plus 0.5em minus 0.4em\relax Hoboken: Wiley,
  2016.

\bibitem{chiou2009application}
C.~Chiou, C.~Chiou, C.~Chu, and S.~Lin, ``The application of fuzzy control on
  energy saving for multi-unit room air-conditioners,'' \emph{Applied thermal
  engineering}, vol.~29, no. 2-3, pp. 310--316, 2009.

\bibitem{ul2014design}
R.~ul~Islam, J.~Iqbal, and Q.~Khan, ``Design and comparison of two control
  strategies for multi-dof articulated robotic arm manipulator,'' \emph{Journal
  of Control Engineering and Applied Informatics}, vol.~16, no.~2, pp. 28--39,
  2014.

\bibitem{nagy2007evaluation}
Z.~K. Nagy, B.~Mahn, R.~Franke, and F.~Allg{\"o}wer, ``Evaluation study of an
  efficient output feedback nonlinear model predictive control for temperature
  tracking in an industrial batch reactor,'' \emph{Control Engineering
  Practice}, vol.~15, no.~7, pp. 839--850, 2007.

\bibitem{kwakernaak1972linear}
H.~Kwakernaak and R.~Sivan, \emph{Linear optimal control systems}.\hskip 1em
  plus 0.5em minus 0.4em\relax Wiley-interscience New York, 1972, vol.~1.

\bibitem{solitine1991}
J.-J.~E. Slotine and W.~Li, \emph{Applied nonlinear control}.\hskip 1em plus
  0.5em minus 0.4em\relax Englewood Cliffs: Prentice Hall, 1991.

\bibitem{sarangapani2006}
J.~Sarangapani, \emph{Neural network control of nonlinear discrete-time
  systems}.\hskip 1em plus 0.5em minus 0.4em\relax Boca Raton: CRC Taylor \&
  Francis, 2006.

\bibitem{Goodfellow2016}
I.~Goodfellow, Y.~Bengio, and A.~Courville, \emph{Deep Learning}.\hskip 1em
  plus 0.5em minus 0.4em\relax MIT Press, 2016,
  \url{http://www.deeplearningbook.org}.

\bibitem{hess2015}
N.~Heess, J.~J. Hunt, T.~P. Lillicrap, and D.~Silver, ``Memory-based control
  with recurrent neural networks,'' \emph{arXiv:1512.04455}, 2015.

\bibitem{jaeger2001}
H.~Jaeger, ``The “echo state” approach to analysing and training recurrent
  neural networks-with an erratum note,'' \emph{Bonn, Germany: German National
  Research Center for Information Technology GMD Technical Report}, vol. 148,
  no.~34, p.~13, 2001.

\bibitem{tanaka2019recent}
G.~Tanaka, T.~Yamane, J.~B. H{\'e}roux, R.~Nakane, N.~Kanazawa, S.~Takeda,
  H.~Numata, D.~Nakano, and A.~Hirose, ``Recent advances in physical reservoir
  computing: a review,'' \emph{Neural Networks}, 2019.

\bibitem{gauthier2018reservoir}
D.~Gauthier, ``Reservoir computing: harnessing a universal dynamical system,''
  \textbf{51}:2, 12 (2018).

\bibitem{lu2017reservoir}
Z.~Lu, J.~Pathak, B.~Hunt, M.~Girvan, R.~Brockett, and E.~Ott, ``Reservoir
  observers: Model-free inference of unmeasured variables in chaotic systems,''
  \emph{Chaos}, vol.~27, no.~4, p. 041102, 2017.

\bibitem{pathak2018model}
J.~Pathak, B.~Hunt, M.~Girvan, Z.~Lu, and E.~Ott, ``Model-free prediction of
  large spatiotemporally chaotic systems from data: a reservoir computing
  approach,'' \emph{Phys. Rev. Lett.}, vol. 120, no.~2, p. 024102, 2018.

\bibitem{vlachas2019}
P.~Vlachas, J.~Pathak, B.~Hunt, T.~P. Sapsis, M.~Girvan, E.~Ott, and
  P.~Koumoutsakos, ``Forecasting of spatial-temporal chaotic dynamics with
  recurrent neural networks: A comparative study of reservoir computing and
  backpropagating algorithms,'' \emph{arXiv:1910.05266}, 2019.

\bibitem{chattopadhyay2019}
A.~Chattopadhyay, P.~Hassanzadeh, and D.~Subramanian, ``Data-driven prediction
  of a multi-scale lorenz 96 chaotic system using deep learning methods:
  Reservoir computing, ann, and rnn-lstm,'' \emph{arXiv:1906.08829}, 2019.

\bibitem{bompas2020}
S.~Bompas, B.~Georgeot, and D.~Gu\'ery-Aodelin, ``Accuracy of neural networks
  for the simulation of chaotic dynamics: precision of training data vs.
  precision of the algorithm,'' \emph{arXiv:2008.04222}, 2020.

\bibitem{jaeger2003adaptive}
H.~Jaeger, ``Adaptive nonlinear system identification with echo state
  networks,'' in \emph{Adv. Neur. In.}, 2003, pp. 609--616.

\bibitem{larger2017high}
L.~Larger, A.~Bayl{\'o}n-Fuentes, R.~Martinenghi, V.~S. Udaltsov, Y.~K. Chembo,
  and M.~Jacquot, ``High-speed photonic reservoir computing using a
  time-delay-based architecture: Million words per second classification,''
  \emph{Phys. Rev. X}, vol.~7, no.~1, p. 011015, 2017.

\bibitem{canaday2018rapid}
D.~Canaday, A.~Griffith, and D.~J. Gauthier, ``Rapid time series prediction
  with a hardware-based reservoir computer,'' \emph{Chaos: An Interdisciplinary
  Journal of Nonlinear Science}, vol.~28, no.~12, p. 123119, 2018.

\bibitem{khodabandehlou2017echo}
H.~Khodabandehlou and M.~S. Fadali, ``Echo state versus wavelet neural
  networks: Comparison and application to nonlinear system identification,''
  \emph{IFAC-PapersOnLine}, vol.~50, no.~1, pp. 2800--2805, 2017.

\bibitem{paraskevopoulos2017modern}
P.~N. Paraskevopoulos, \emph{Modern control engineering}.\hskip 1em plus 0.5em
  minus 0.4em\relax CRC Press, 2017.

\bibitem{antonik2016}
P.~Antonik, M.~Hermans, F.~Duport, M.~Haelterman, and S.~Massar, ``Towards
  pattern generation and chaotic series prediction with photonic reservoir
  computers,'' in \emph{SPIE LASE}.\hskip 1em plus 0.5em minus 0.4em\relax
  International Society for Optics and Photonics, 2016, pp. 97\,320B--97\,320B.

\bibitem{jaeger2001echo}
H.~Jaeger, ``The “echo state” approach to analysing and training recurrent
  neural networks-with an erratum note,'' \emph{Bonn, Germany: German National
  Research Center for Information Technology GMD Technical Report}, vol. 148,
  no.~34, p.~13, 2001.

\bibitem{gallicchio2017deep}
C.~Gallicchio, A.~Micheli, and L.~Pedrelli, ``Deep reservoir computing: A
  critical experimental analysis,'' \emph{Neurocomputing}, vol. 268, pp.
  87--99, 2017.

\bibitem{takens1981detecting}
F.~Takens, ``Detecting strange attractors in turbulence,'' in \emph{Dynamical
  systems and turbulence, Warwick 1980}.\hskip 1em plus 0.5em minus 0.4em\relax
  Springer, 1981, pp. 366--381.

\bibitem{lu2018}
Z.~Lu, B.~R. Hunt, and E.~Ott, ``Attractor reconstruction by machine
  learning,'' \emph{Chaos}, vol.~28, p. 061104, 2018.

\bibitem{lukovsevivcius2012practical}
M.~Luko{\v{s}}evi{\v{c}}ius, ``A practical guide to applying echo state
  networks,'' in \emph{Neural networks: Tricks of the trade}.\hskip 1em plus
  0.5em minus 0.4em\relax Berlin, Heidelberg: Springer, 2012, pp. 659--686.

\bibitem{griffith2019forecasting}
A.~Griffith, A.~Pomerance, and D.~J. Gauthier, ``Forecasting chaotic systems
  with very low connectivity reservoir computers,'' \emph{arXiv preprint
  arXiv:1910.00659}, 2019.

\bibitem{norgaard2000neural}
P.~M. N{\o}rg{\aa}rd, O.~Ravn, N.~K. Poulsen, and L.~K. Hansen, ``Neural
  networks for modelling and control of dynamic systems-a practitioner's
  handbook,'' 2000.

\bibitem{rivera2003plant}
D.~E. Rivera, H.~Lee, M.~W. Braun, and H.~D. Mittelmann, ``" plant-friendly"
  system identification: a challenge for the process industries,'' \emph{IFAC
  Proceedings Volumes}, vol.~36, no.~16, pp. 891--896, 2003.

\bibitem{waegeman2012feedback}
T.~Waegeman, B.~Schrauwen \emph{et~al.}, ``Feedback control by online learning
  an inverse model,'' \emph{IEEE T. Neur. Net. Lear.}, vol.~23, no.~10, pp.
  1637--1648, 2012.

\bibitem{Shinbrot1993}
T.~Shinbrot, C.~Grebogi, J.~A. Yorke, and E.~Ott, ``Using small perturbations
  to control chaos,'' \emph{Nature (London)}, vol. 363, pp. 411--417, 1993.

\bibitem{chang1998stabilizing}
A.~Chang, J.~C. Bienfang, G.~M. Hall, J.~R. Gardner, and D.~J. Gauthier,
  ``Stabilizing unstable steady states using extended time-delay
  autosynchronization,'' \emph{Chaos}, vol.~8, no.~4, pp. 782--790, 1998.

\end{thebibliography}

\begin{IEEEbiography}[{\includegraphics[width=1in,height=1.25in,clip,keepaspectratio]{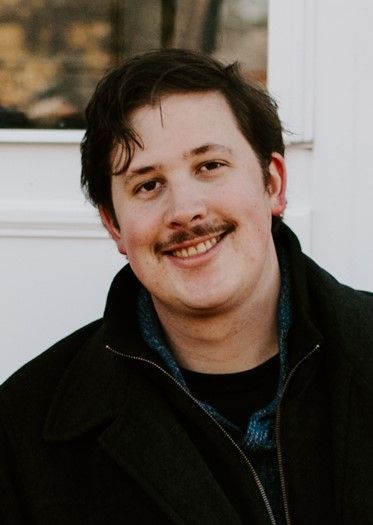}}]{Daniel Canaday}
was born in Columbus, OH in 1991. He received the B.S. degree in physics and mathematics from Ohio State University, Columbus, OH, USA in 2014, the M.S. degree in physics from Ohio State University, Columbus, OH, USA in 2017, and the Ph.D. degree in physics from Ohio State University, Columbus, OH, USA in 2019.

He is currently a scientist at Potomac Research, LLC, Alexandria, VA. His research is concerned with applied reservoir computing and the application of physical neural networks to cryptography.
\end{IEEEbiography}

\begin{IEEEbiography}[{\includegraphics[width=1in,height=1.25in,clip,keepaspectratio]{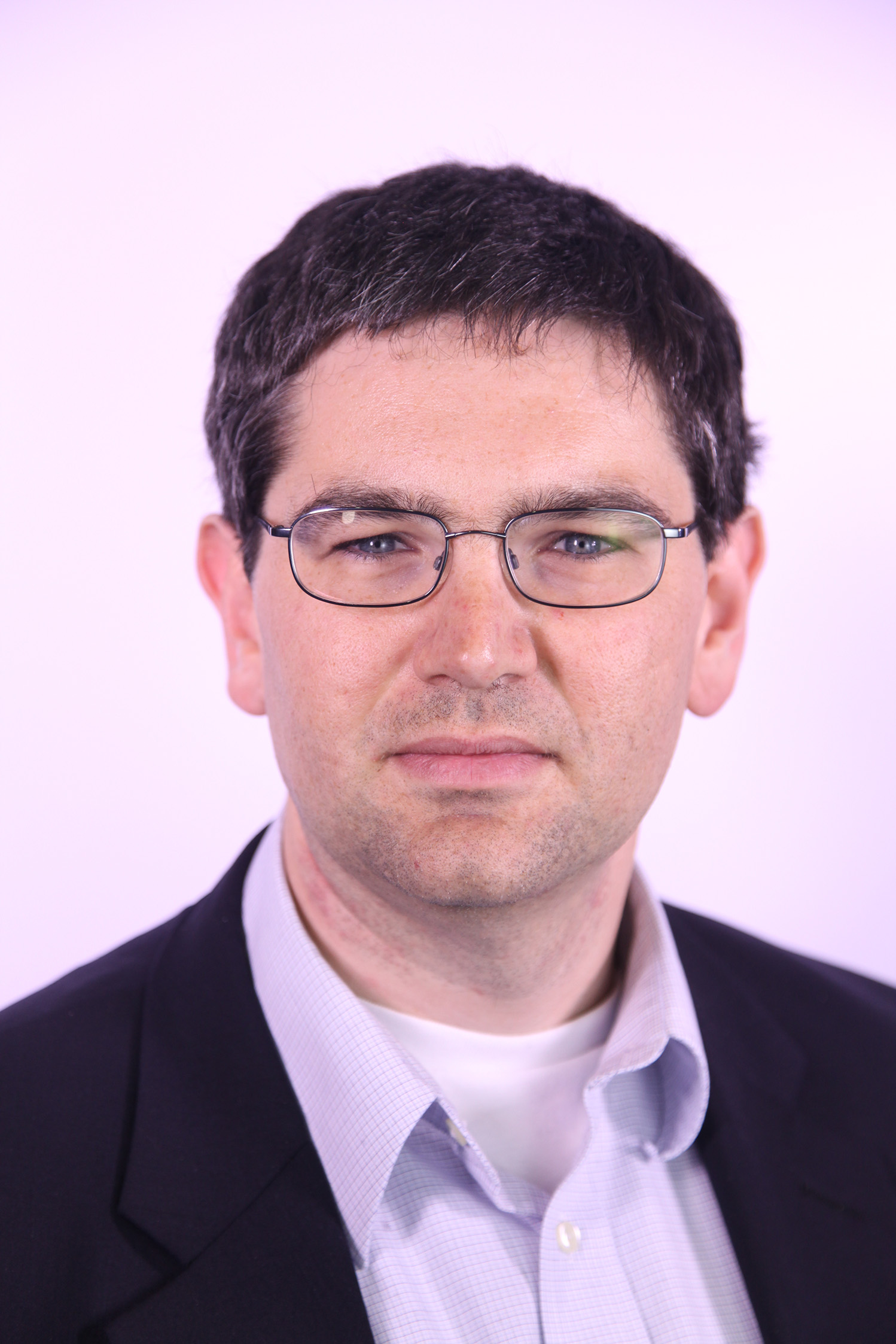}}]{Andrew Pomerance}
Andrew Pomerance  was born in Washington, DC, USA, in 1980. He received the B.S. and M.S. degrees in Electrical and Computer Engineering from Carnegie Mellon University, Pittsburgh, PA, in 2002, and the Ph.D. in Physics from the University of Maryland, College Park, MD, in 2009. 

From 2009 to 2013, he was with Raytheon Applied Signal Technology, Tyson's Corner, VA, USA. He is currently the president of Potomac Research, LLC, Alexandria, VA, USA.  His research is concerned with nonlinear dynamics with applications to machine learning and cryptography.
\end{IEEEbiography}

\begin{IEEEbiography}[{\includegraphics[width=1in,height=1.25in,clip,keepaspectratio]{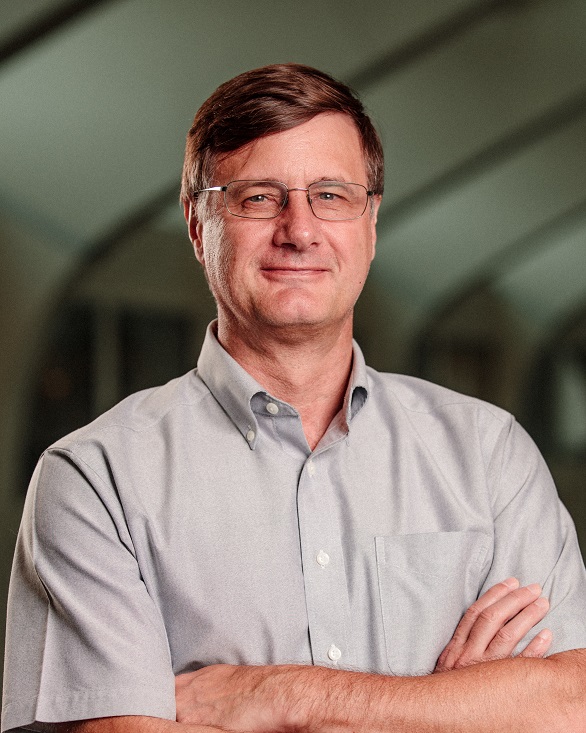}}]{Daniel J. Gauthier}
Daniel J. Gauthier is a Professor of Physics and Electrical and Computer Engineering at The Ohio State University. He received the B.S., M.S., and Ph.D. degrees from the University of Rochester, Rochester, NY, in 1982, 1983, and 1989, respectively. His Ph.D. research on “Instabilities and chaos of laser beams propagating through nonlinear optical media” was supervised by Prof. R.W. Boyd and supported in part through a University Research Initiative Fellowship. From 1989 to 1991, he developed the first CW two-photon optical laser as a Post-Doctoral Research Associate under the mentorship of Prof. T.W. Mossberg at the University of Oregon. In 1991, he joined the faculty of Duke University, Durham, NC, as an Assistant Professor of Physics and was named a Young Investigator of the U.S. Army Research Office in 1992 and the National Science Foundation in 1993.  He was the Robert C. Richardson Professor of Physics at Duke from 2011- 2015, chair of the Duke Physics Department from 2005 – 2011, interim chair in spring 2015, and was a founding member of the Duke Fitzpatrick Institute for Photonics. He moved to The Ohio State University in 2016. His research interests include: reservoir computing, synchronization and control of the dynamics of complex networks in electronic and optical systems, quantum communication, and nonlinear quantum optics. Prof. Gauthier is a Fellow of the Optical Society of America and the American Physical Society.
\end{IEEEbiography}

\end{document}